Title Page

Title: An Exploratory Data Survey of Drug Name Incidence and Prevalence From the FDA's Adverse Event Reporting System, 2004-2012Q2

Running Head: Drug Names Surveillance and the FDA Adverse Event Reporting System

Authors: Nick Williams

Institutions: New York University Langone Medical Center

Corresponding Author: Nick Williams

Name: Nick Williams
Address:
*NYU School of Medicine, Department of Emergency Medicine*
*Bellevue Hospital Center*
*462 First Avenue, OBV A345, New York, NY 10016*
Telephone: 1-347-327-8684
Fax: *212.562.3001*
Email: nick.williams@nyumc.org

Key Points
1. Historical reporting trends for AE are relatively easy to produce from the publicly available version of FAERS.
2. Data mining approaches should mirror established epidemiology population level surveillance operations.
3. Specificity aside, there is much to learn from a maximum sensitivity analysis of drug names in an AE database.
4. Pharmacoepidemiology efforts should be driven by epidemiology surveillance of population level values to maximize public health benefits for the largest segment of the affected population.

Conflict of Interest: The author has no competing interests to declare. This survey is unfunded.

Word Count: 2510

Prior Postings and Publication: The author declares no prior presentation of this work of any kind or in any format.

Abstract Title:

An Exploratory Data Survey of Drug Name Incidence and Prevalence From the FDA's Adverse Event Reporting System, 2004-2012Q2

Purpose:
To count and monitor the drug names reported in the publicly available version of the Federal Adverse Event Reporting System (FAERS) from 2004-2012Q2 in a maximized sensitivity relational model.

Methods:
Data mining and data modeling was conducted and event based summary statistics with plots were created from nine continuous years of FAERS data.

Results:
This FAERS model contains 344,452 individual drug names and 432,541,994 drug name count references which occurred across 4,148,761 human subjects in the 34 quarter study period. FAERS has several trending outbreaks of drug name incidence reported for Adverse Events (AE). Plots for the top 100 scoring drug name references are reported by year and quarter; the top 100 drug names contain 143,384,240 references or 33% of all drug name references over 34 quarters of continuous FAERS data.

Conclusions:
While FAERS contains many drugs and adverse event reports, its data pertains to very few of them. Drug name incidence lends timely and effective surveillance of large populations of Averse Event Reports and does not require the cause of the AE, nor its validity to be known to detect a mass poisoning. Drug name surveillance and incidence reporting may serve as viable alternative to odds ratio's and other Gaussian based statistical approaches when a maximized sensitivity relational model is used.

Introduction

The Federal Adverse Event Reporting System (FAERS) data set is a massive, publicly available pharmacoepidemiology repository for worldwide post marketing drug surveillance[i]. FAERS is populated with provider reported data points describing Adverse Events (AE) experienced by patients who consume marketed pharmaceutical drugs in patient care settings. As more and more drugs are marketed over time and access to western pharmaceuticals continues to grow it stands to reason that we will see more adverse events. How many adverse events is too many[ii] for a single drug name is a question of active debate, with obvious cases like (contaminated) Heparin Sodium Injection (HSI)[iii] and VIOXX[iv] on one hand and well debated yet indecisive cases like Aspirin[v] on the other. While these drug names are often explored in FAERS in terms of historical trends[vi] of individual drug names, odds ratio signal dis-proportionality and mortality surveillance[vii], the literature fundamentally lacks summary statistics for all drug names in FAERS or a drug name to drug names comparison.

Here, a maximum sensitivity model is crafted from FAERS reports from 2004-2012Q2; utilizing the FDA's publicly available database. This data source is largely discarded as a research resource because FAERS contains over 300,000 drug names for a highly likely 10,000 substances[viii]. This inflation is largely due to spelling errors and open input data string fields. Further there is widespread criticism that FAERS does not contain meaningful data in the public version and that most impressive data elements are reserved for government investigators in the name of patient information protections[ix]. These concerns are accurate, but must be tempered by efforts to develop surveillance methodologies that can resolve these static roadblocks that have shown no sign of moving despite years of publicly available surveillance data.

FAERS uses a three tier index model where drug names are tied to clinical indications and observed reactions by the reporting provider. A patient may have several drugs (poly pharmacy[x]), indications (co-morbidity) and reactions (adverse or clinical, known or unknown) in any single subject level report. A host of secondary variables, including patient outcome is also available. Several common statistical and epidemiological methods of mining FAERS may well be inappropriate given the structure, distribution and shape of FAERS drug name incidence and prevalence.

For a direct example, a patient on ten drugs with four indications and two reactions returns six drug name counts across each of the ten drugs for the individual subject in this model, inflated by indication and reaction. To become a high count relational drug name this must happen to a specific drug disproportionately across bodies or time. Although this model sacrifices the clinical sensitivity that many providers and toxicologists look to FAERS to provide, it allows epidemiologists to describe in explicit detail the incidence and relevance of reported drug name relationships over time. More complex FAERS reports suggest more complex management of and therefore spontaneous or unknown reactions and severe subject level clinical complications. Population level events outside of bedside matters are well suited to maximum sensitivity detection methods like this.

Knowing the noise from the pharmacokinetics is a major undertaking of data science and pharmacology. Signal based reporting was supposed to solve this problem[xi], yet signal work is largely derived from single drug odds ratios[xii] [xiii]that assume proportional incidence can be subject to false positives and noise[xiv]. By assuming FAERS reports are false until proven true (as if FAERS reports were populated by clinical providers by accident[xv]) we obscure and deliberately under power FAERS signals with overly complicated mathematical models. Different and multiple approaches to AE surveillance may resolve longstanding dissatisfaction with championed single method approaches.

Methods

Subject numbers (ISR) were left joined to their reported drug names, then clinical indications and finally reactions. This set was then striped of its subject level identifiers in Microsoft Access 2008. Further, clinical indication and specific reaction were also striped from the model to create a relational count data model of drug names by quarter using Google API Big-Query. This model assumes every reported drug name in FAERS actually influenced an adverse event and that in a population level perspective, higher count values indicate more problematic substances. While FAERS probably contains false positives, they most likely did not happen across the available, international and historical patient population contained in FAERS. Plots were constructed using Microsoft Excel 2007, RED-R (R programming language[xvi]) and CIRCOS[xvii]. Summary statistics were computed in SPSS 19.

Results

Table One: Model Values

| Model Event | Model Event | Drug Name | Drug Name |
|---|---|---|---|
| Sum | 432,541,994 | Drug Names | 344,452 |
| Mean | 1,255.739534 | Minimum Count | 1 |
| Std Error of Mean | 61.42018455 | Maximum Count | 12,673,689 |
| Std Deviation | 36,047.52672 | Count Median | 10 |
| Variance | 12,99424182 | Count Mode | 1 |
| Skewness | 162.2495311 | Percentiles | Count |
| Std Error of Skewness | 0.004173586 | 25Th | 4 |
| Kurtosis | 47,542.17632 | 50Th | 10 |
| Std Error of Kurtosis | 0.008347148 | 75Th | 40 |
| Range | 12,673,688 | | |

Over 75% of the 344,452 drug names in FAERS contained less than forty relational counts across 4,148,761 human subjects and 34 reporting quarters. Despite maximum sensitivity most drug names failed to capture a meaningful volume of references over time suggesting non-population level AE but individual patient AE. Some drug names returned millions of counts suggesting mass poisonings.

Table Two: Quarterly (Q) Measures for The Top Ten Scoring Drug Names

| DRUG NAME | QSUM | QMIN | QMAX | QMEDIAN | QAVERAGE | QSD |
|---|---|---|---|---|---|---|
| HEPARIN SODIUM INJECTION | 12,673,689.00 | 4.00 | 3,201,998.00 | 2,772.50 | 396,052.78 | 784,504.60 |
| ASPIRIN | 5,023,238.00 | 44,465.00 | 457,536.00 | 86,394.00 | 147,742.29 | 114,208.94 |
| FOSAMAX | 4,762,966.00 | 6,352.00 | 699,062.00 | 75,132.50 | 140,087.24 | 176,753.36 |
| HUMIRA | 3,258,678.00 | 4,155.00 | 487,234.00 | 76,279.50 | 95,843.47 | 102,031.13 |
| SEROQUEL | 3,149,210.00 | 7,841.00 | 393,928.00 | 26,553.00 | 92,623.82 | 111,237.75 |
| VIOXX | 3,064,082.00 | 2,119.00 | 455,405.00 | 18,932.00 | 90,120.06 | 130,925.94 |
| PREDNISONE | 2,749,852.00 | 3,790.00 | 298,614.00 | 43,194.50 | 80,878.00 | 78,592.22 |
| LASIX | 2,682,382.00 | 23,422.00 | 291,380.00 | 43,609.00 | 78,893.59 | 67,108.05 |
| METHOTREXATE | 2,288,678.00 | 18,771.00 | 201,030.00 | 43,500.00 | 67,314.06 | 51,855.42 |
| LIPITOR | 2,097,675.00 | 18,981.00 | 182,054.00 | 40,572.00 | 61,696.32 | 43,485.15 |
| LISINOPRIL | 2,042,128.00 | 6,701.00 | 217,668.00 | 27,557.50 | 60,062.59 | 57,989.02 |

HSI is the largest scoring drug name in the model followed by the widely used and debated Aspirin. The distinction between their QSUM is telling as HSI is nearly three times larger than Aspirin. Several drugs beat out VIOXX for the third spot on the list and warrant further investigation. The range between the QMIN and QMAX may prove adequate to detect departures from the norm if taken with median and average taken as baseline values. All values are quarterly except QSUM. Mass poisoning events are detectable when historical subservience is utilized.

Graph One: Box Plot of Drug Name Reference Counts by Quarter

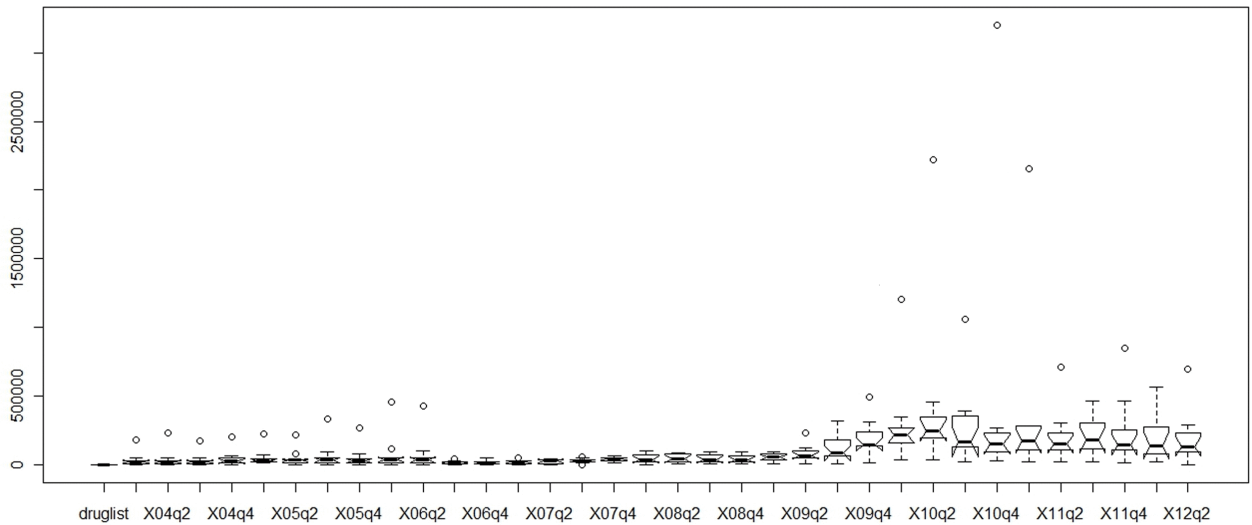

VIOXX (Graph 1 Box Plot Left) and HEPERIN SODIUM INJECTION (Graph 1 Box Plot Right) are clearly legible and served for the top scoring values over several quarters of documented historical incidence. This data model can clearly detect departure from the trend with simplistic incidence survey work.

Graph Two: Population Level Ratio Trend Variation in FAERS

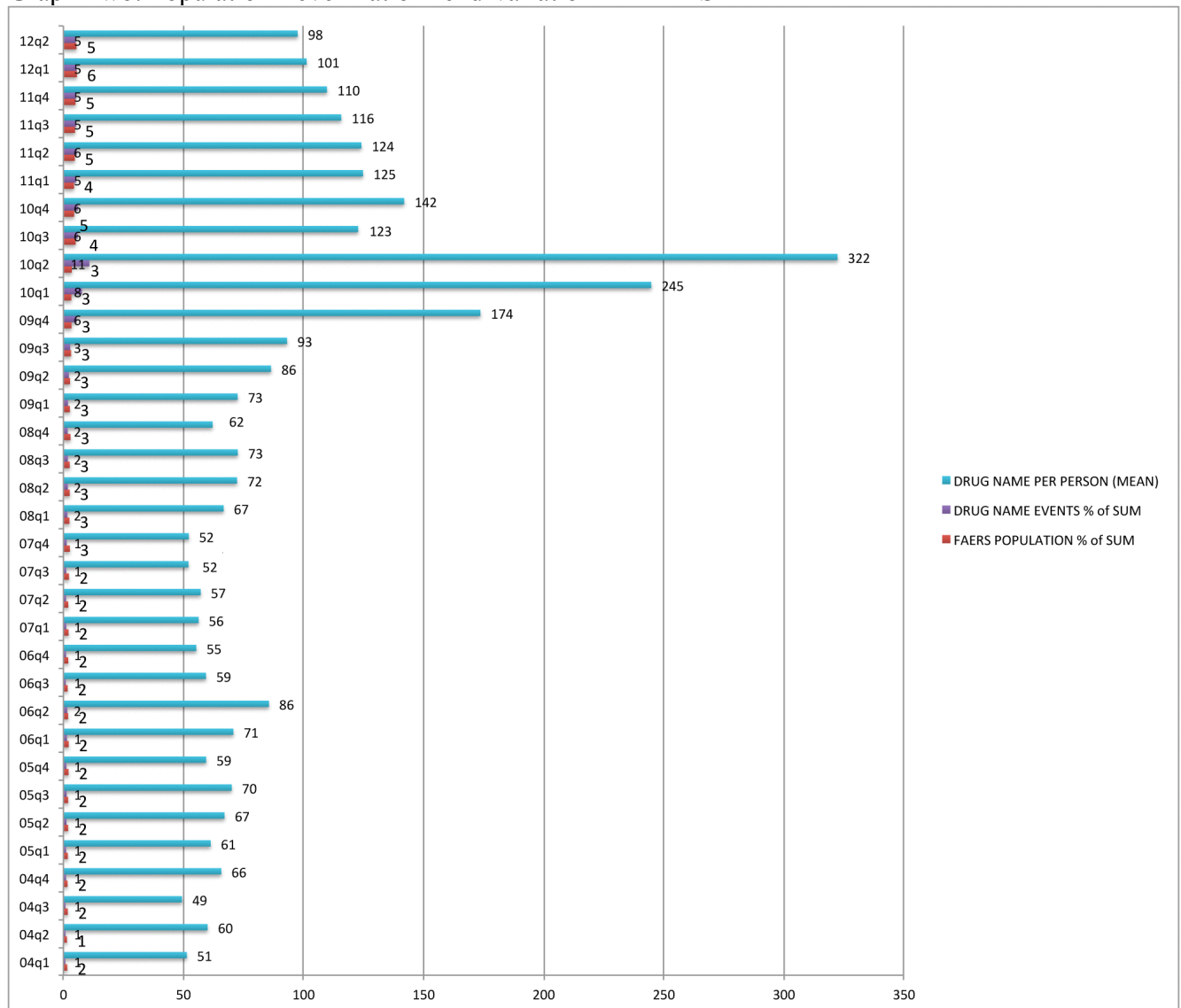

Here we see the total population from the study period by quarter divided by the number of drug name reference events by quarter and plotted against the percentage of reporting subjects and drug name reference events from the study period. We clearly see that while there is no stable rate or neutral state in FAERS there are strong departures from the norm, especially coinciding with HSI contamination. VIOXX and other poisoning events are not readily legible here, suggesting that some mass poisonings are obscured by proportional surveillance. If FAERS is a natural pattern without variation, the affected human population and drug name counts should return similar percentages over time. There are several departures from the expected 1:2 ratio, where the complexity of the adverse event outpaced the human population experiencing it.

Top 100 Sum Drug Name References Set Plotted by Year and Quarter 2004-2012

This set contains 100 drug names and 33% of FAERS drug name count references over nine continuous years.

2004

2005

2006

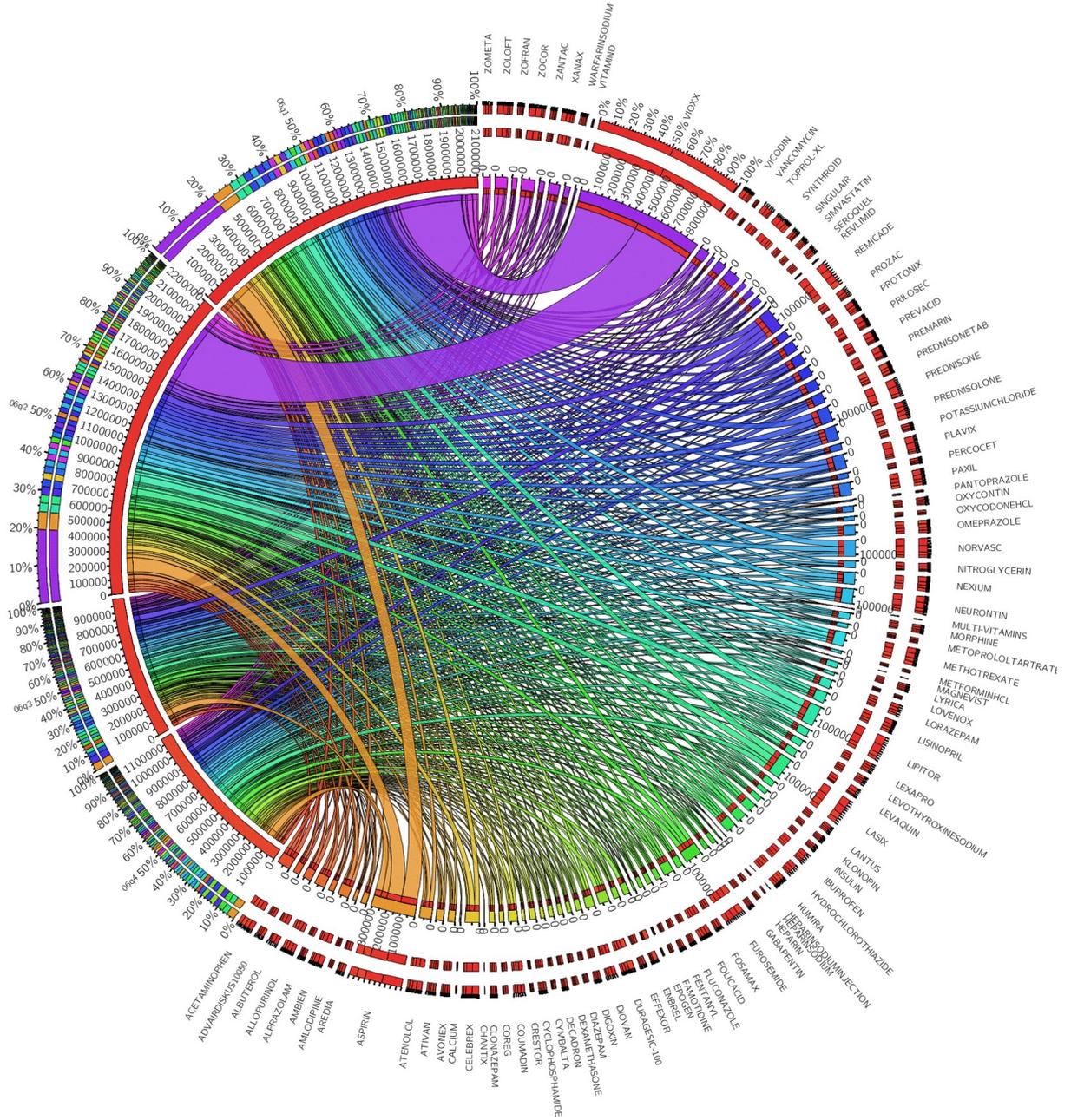

2007

2008

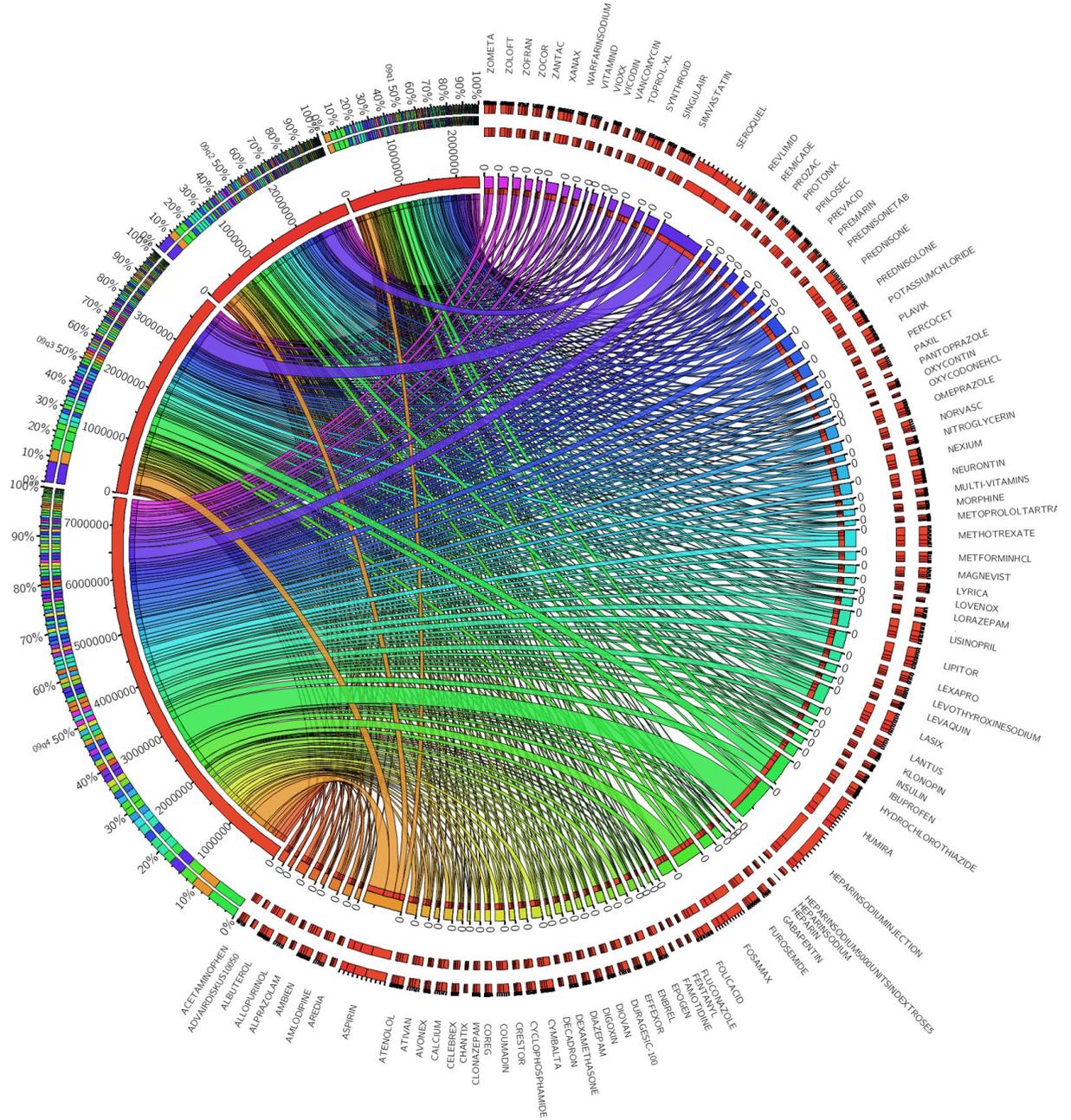

2009

2010

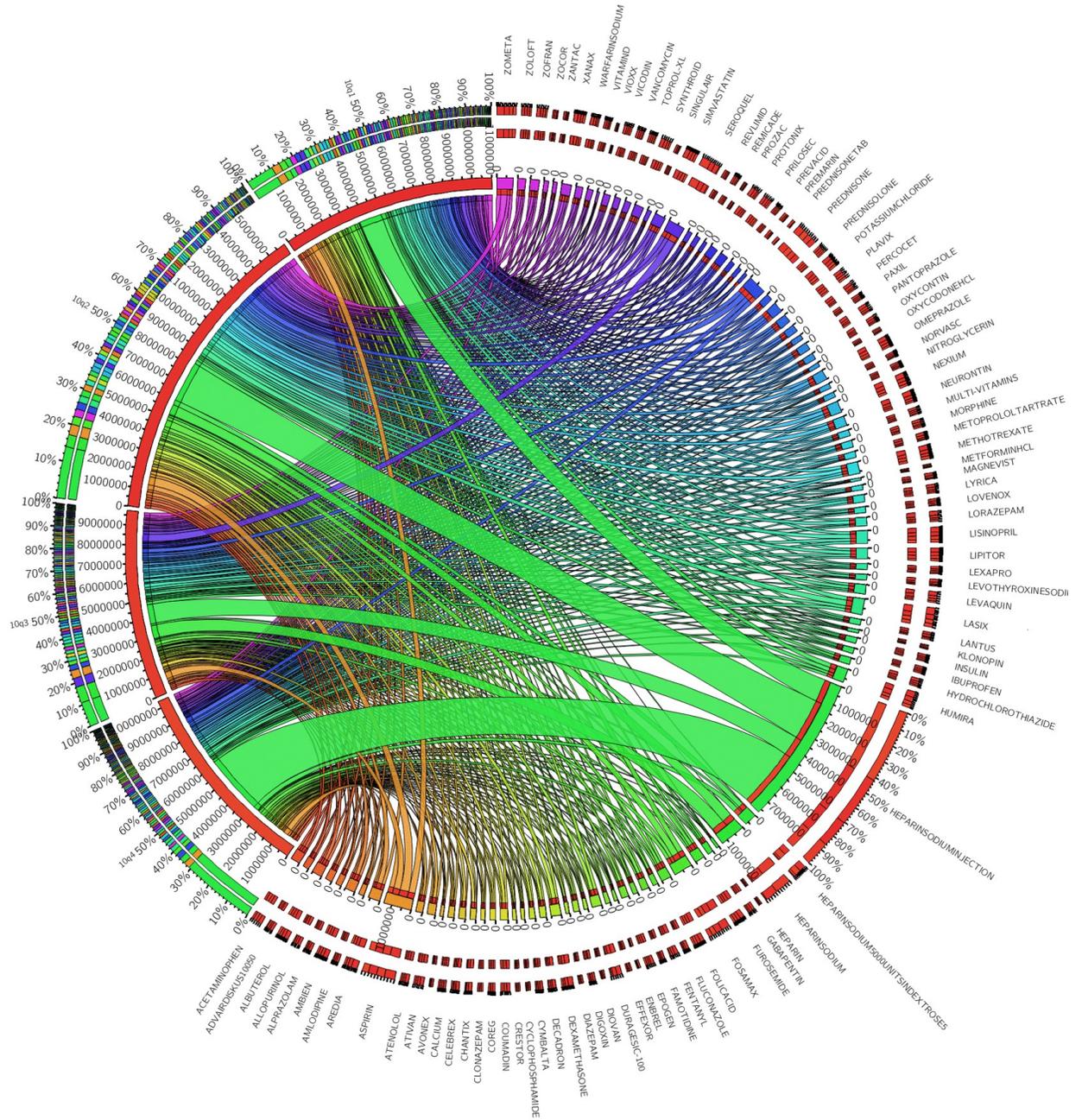

2011

2012

Full Set 2004-2012Q2

Discussion:

There is an order of operations to population level epidemiology: sensitivity, outbreak identification, specificity, case definition and then automated surveillance. This maximum sensitivity model may be complicated to understand as it has no specificity controls. In the first step specificity or cause is not required, rather detection is key. These AE case definitions emerge from drilling down into initial sensitivity surveillance. This model can detect signals of AE despite the equal power that relational modeling lends to assumed false positives. Further, historical epidemics of AE (VIOXX and HSI) are detectable.

Most importantly this model does not rely on any base measure of a drug over time but rather compares a drug to drugs across clinical complexity within and across quarters. VIOXX may be a successful yet late intervention; as most surveillance schemes look for increases in reporting for a baseline that was never natural. Further, online supplement Graph FAERS 2006 clearly demonstrates the power and utility of FAERS by highlighting the near evaporation of VIOXX cases by the second quarter of 2006; for which mass legal action[xviii] rather than marketing practices may well be responsible. Further utility can be seen in countless label changes, black box warnings and direct actions taken by the FDA. Increasingly FAERS has fallen under somewhat unwarranted criticism for not detailing epidemics or naming drugs as endemic causes of AE fast enough. This model details the kinds of utility that may be found in an adverse event reporting system like FAERS and suggest drug names for further investigation. Maximum sensitivity seems to coincide with major epidemics of AE including VIOXX and HSI, suggesting that other model values may also be valid. High scoring drugs warrant further epidemiological investigation, as these signals are equally powered in this maximum sensitivity model as VIOXX and HSI.

Conclusion

FAERS presents challenges and rewards in an endless waltz of suspicion, corroboration and false positives in traditional surveillance scheme. Novel approaches as well as model, relational and dimensional analysis may supplement formal surveillance programs. While more invention, collaboration and expertise are often called upon to supplement surveillance efforts, an old fashioned closed set count data model has demonstrated some utility.

i US Food and Drug Administration (12/01/2013) 'FDA Adverse Event Reporting System **(FAERS) (formerly AERS)' Last** Updated:09/10/2012; Retrieved from: http://www.fda.gov/Drugs/GuidanceComplianceRegulatoryInformation/Surveillance/AdverseDrugEffects/default.htm

ii Lester, J., Neyarapally, G. A., Lipowski, E., Graham, C. F., Hall, M. and Dal Pan, G. (2013), Evaluation of FDA safety-related drug label changes in 2010. Pharmacoepidem. Drug Safe., 22: 302–305. doi: 10.1002/pds.3395

iiiRodriguez, E. M., Staffa, J. A. and Graham, D. J. (2001), The role of databases in drug postmarketing surveillance. Pharmacoepidem. Drug Safe., 10: 407–410. doi: 10.1002/pds.615

iv Eric J. Topol, M.D. Failing the Public Health — Rofecoxib, Merck, and the FDA N Engl J Med 2004; 351:1707-1709October 21, 2004 DOI: 10.1056/NEJMp048286

v Kenneth R. McQuaid, MD, Loren Laine, MD Systematic Review and Meta-analysis of Adverse Events of Low-dose Aspirin and Clopidogrel in Randomized Controlled Trials The American Journal of Medicine, Volume 119, Issue 8, August 2006, Pages 624–638 Http://dx.doi.org/10.1016/j.amjmed.2005.10.039

vi AM Hochberg[1] and M Hauben Time-to-Signal Comparison for Drug Safety Data-Mining Algorithms vs. Traditional Signaling Criteria Clinical Pharmacology & Therapeutics (2009); 85, 6, 600–606 doi:10.1038/clpt.2009.26

vii Jennifer Jacobs, Peter Fisher Polypharmacy, multimorbidity and the value of integrative medicine in public health European Journal of Integrative Medicine, Volume 5, Issue 1, February 2013, Pages 4–7 http://dx.doi.org/10.1016/j.eujim.2012.09.001

viii Bilker, W., Gogolak, V., Goldsmith, D., Hauben, M., Herrera, G., Hochberg, A., Jolley, S., Kulldorff, M., Madigan, D., Nelson, R., Shapiro, A. and Shmueli, G. (2006), Accelerating statistical research in drug safety. Pharmacoepidem. Drug Safe., 15: 687–688. doi: 10.1002/pds.1267

ix Bilker, W., Gogolak, V., Goldsmith, D., Hauben, M., Herrera, G., Hochberg, A., Jolley, S., Kulldorff, M., Madigan, D., Nelson, R., Shapiro, A. and Shmueli, G. (2006), Accelerating statistical research in drug safety. Pharmacoepidem. Drug Safe., 15: 687–688. doi: 10.1002/pds.1267

x Raymond L. Woosley MD, PhD Discovering adverse reactions: Why does it take so long? Clinical Pharmacology & Therapeutics (2004) 76, 287–289; doi: 10.1016/j.clpt.2004.06.006

xi Stephenson, W. P. and Hauben, M. (2007), Data mining for signals in spontaneous reporting databases: proceed with caution. Pharmacoepidem. Drug Safe., 16: 359–365. doi: 10.1002/pds.1323

xii Poluzzi, E., Raschi, E., Moretti, U. and De Ponti, F. (2009), Drug-induced **torsades de pointes**: data mining of the public version of the FDA Adverse Event Reporting System (AERS). Pharmacoepidem. Drug Safe., 18: 512–518. doi: 10.1002/pds.1746

xiii Almenoff, J. S., DuMouchel, W., Kindman, L. A., Yang, X. and Fram, D. (2003), Disproportionality analysis using empirical Bayes data mining: a tool for the evaluation of drug interactions in the post-marketing setting. Pharmacoepidem. Drug Safe., 12: 517–521. doi: 10.1002/pds.885

xiv Brown, J. S., Kulldorff, M., Petronis, K. R., Reynolds, R., Chan, K. A., Davis, R. L., Graham, D., Andrade, S. E., Raebel, M. A., Herrinton, L., Roblin, D., Boudreau, D., Smith, D., Gurwitz, J. H., Gunter, M. J. and Platt, R. (2009), Early adverse drug event signal detection within population-based health networks using sequential methods: key methodologic considerations. Pharmacoepidem. Drug Safe., 18: 226–234. doi: 10.1002/pds.1706

xv [Tannert C](), [Elvers HD](), [Jandrig B]().The ethics of uncertainty. In the light of possible dangers, research becomes a moral duty. [EMBO Rep.]() 2007 Oct;8(10):892-6. PMID:17906667 [PubMed - indexed for MEDLINE] PMCID: PMC2002561

xvi Covington, K. R. and A. Parikh (2011, August). The red-r framework for integrated discovery. The Red-R Journal 1-08/08/2011.

xviiKrzywinski M, Schein J, Birol I, et al. Circos: an information aesthetic for comparative genomics. Genome Research. 2009;19:1639–1645.

xviii Thomas, W. John Vioxx Story: Would It Have Ended Differently in the European Union, The; 32 Am. J.L. & Med. 366 (2006)